# Antiferromagnetism in Ru$_2$MnZ (Z=Sn, Sb, Ge, Si) full Heusler alloys: effects of magnetic frustration and chemical disorder

Sergii Khmelevskyi[1], Eszter Simon[1] and László Szunyogh[1,2]

[1] Department of Theoretical Physics, Budapest University of Technology and Economics, Budafoki út 8., H-1111 Budapest, Hungary

[2] MTA-BME Condensed Matter Research Group, Budapest University of Technology and Economics, Budafoki út 8., H-1111 Budapest, Hungary

We present systematic theoretical investigations to explore the microscopic mechanisms leading to the formation of antiferromagnetism in Ru$_2$MnZ (Z= Sn,Sb,Ge,Si) full Heusler alloys. Our study is based on first-principles calculations of inter-atomic Mn-Mn exchange interactions to set up a suitable Heisenberg spin-model and on subsequent Monte-Carlo simulations of the magnetic properties at finite temperature. The exchange interactions are derived from the paramagnetic state, while a realistic account of long-range chemical disorder is made in the framework of the Coherent Potential Approximation. We find that in case of the highly ordered alloys (Z=Sn and Sb) the exchange interactions derived from the perfectly ordered L2$_1$ structure lead to Néel temperatures in excellent agreement with the experiments, whereas, in particular, in case of Si the consideration of chemical disorder is essential to reproduce the experimental Néel temperatures. Our numerical results suggest that improving a heat treatment of the samples to suppress the intermixing between the Mn and Si atoms, the Néel temperature of the Si-based alloys can potentially be increased by more than 30%. Furthermore, we show that in strongly disordered Ru$_2$MnSi alloys a distinct change in the antiferromagnetic ordering occurs.

## I. Introduction.

Because of their challenging magnetic properties Heusler alloys have attracted considerable attention in the last decades[1,2,3]. The possibility of tuning the magnetic and electronic properties by varying the alloy composition to a large extent, while keeping the crystal structure unchanged, provided an opportunity to verify various fundamental theoretical concepts related to magnetic alloys (see e.g. Ref. [4] and references therein). The main reason, which brought ferromagnetic Heusler alloys to the frontline of research, is related to the possibility of full spin-polarization at the Fermi level due to the half-metallic character of their electronic structure[5]. This feature is believed to have an essential impact on developing highly efficient spintronics devices[6].

Most of the magnetic Heusler alloys exhibit local magnetic moments[7] that can successfully be described in the framework of the Heisenberg spin model. It is therefore no surprise that the exchange interactions in Heusler alloys have been extensively studied on ab-initio level[4]. Due to their possible application in magnetic shape memory devices, ferromagnetic Mn based alloys with the chemical formula $X_2MnZ$, where X is a transition metal element and Z is a *p*-element, gained a particular interest[4,8,9,10,11,12]. It was widely revealed that the first nearest neighbor magnetic interactions between the Mn atoms in the $L2_1$ full Heusler alloy crystal structure (see upper panel of Fig. 1), is strongly ferromagnetic and adds the main contribution to the Curie temperature. However, if , due to an excess of Mn in the alloy composition or due to chemical disorder in stoichiometric samples, Mn atoms are also present on the Z (*p*-element) sites, they might interact antiferromagnetically with the nearest neighbor (NN) Mn atoms on the proper (or original) sites and the system might become ferrimagnetic. In particular, this is the case for the $Ni_2MnAl$ alloy, where a high degree of chemical disorder can be achieved by suitable thermal treatment. This material can even be a compensated antiferromagnet in the disordered B2 phase[13] which is called structurally induced antiferromagnetism[12].

The growing interest in new metallic antiferromagnets, triggered by their application in spintronics devices[14,15], focuses the attention also to antiferromagnetic (AFM) Heusler alloys[16,17]. However, only relatively few antiferromagnetic Heusler alloys are known with sufficiently high Néel temperature ($T_N$)1, which might raise some doubts against their application in technology. In the $Ru_2MnZ$ (Z = Sn, Sb, Ge, Si) alloys the Mn moments order on the four fcc sublattices into the highly frustrated, so-called 2$^{nd}$ kind of AFM structure (see lower panel of Fig. 1). The Néel temperatures of the $Ru_2MnSi$ and $Ru_2MnGe$ compounds are slightly above room temperature[18] (313 K and 316 K, respectively), and in contrast to the relatively high temperature AFM $Ni_2MnAl$ alloy, they are proper antiferromagnets in the fully ordered state. $Ru_2MnSn$ has somewhat lower ordering temperature ($T_N$ =296 K) but still slightly above the room temperature, while $Ru_2MnSb$ has the lowest Néel temperature (195 K) in this series[18]. Recently, some attempts have been made to increase the Néel temperatures of these compounds by producing strained epitaxial films[17].

The quite low critical temperatures of the known $X_2$MnZ AFM Heusler alloys have at least two main reasons. The first reason is the ferromagnetic character of the 1$^{st}$ NN Mn-Mn coupling found in first principles calculations. In Ref. [19] the first three NN exchange couplings in the $Ru_2$MnZ series have been estimated from total energy calculations for a few ordered magnetic configurations. It has been concluded that the 1$^{st}$ NN interaction is ferromagnetic and the stabilization of the AFM structure occurs due to almost equally strong 2$^{nd}$ NN exchange coupling of AFM character. Another source of the low critical temperatures of AFM Heusler alloys is the magnetic frustration that can naturally occur on an fcc lattice. In particular, the $Ru_2$MnZ compounds exhibit a very special and rare type of AFM order on fcc lattice, the so-called 2$^{nd}$ kind of AFM structure[18]. As visualized in the lower panel of Fig. 1, in this type of ordering each of the four simple cubic lattices constituting the fcc lattice have a checkerboard AFM order, but the mutual inter-sublattice orientation of the magnetic moments is completely frustrated. Within the Heisenberg model this frustration can only be resolved by quantum effects and it has been subject of a number of theoretical investigations[20,21].

In this work we present a thorough first-principles study of antiferromagnetism in the $Ru_2$MnZ series of Heusler alloy. We pay particular attention to the effects of chemical and magnetic disorder on the calculated exchange constants and on the magnetic frustration. The influence of disorder on the magnetic transition temperature in Heusler alloys was studied on a similar level of accuracy only in Ref. [22] in case of the half-Heusler NiMnSb alloy, where the importance of the non mean-field treatment of the chemical disorder within a Heisenberg model has been pointed out. This obviously applies to the $Ru_2$MnZ full Heusler alloys with non-trivial AFM ordering. The magnetic features related to the special AFM ordering will be discussed for the case of the $Ru_2$MnSb compound in details. We found that in case of Z = Si the experimentally observed chemical anti-site disorder on Mn and Si sublattices considerably reduces the Néel temperature, opening the way for a sample improvement in terms of a suitable heat treatment. In case of strongly disordered $Ru_2$MnSi compound, close to the disordered B2 phase, our study predicts a transition to a complex AFM structure essentially different from the 2$^{nd}$ kind of AFM ordering.

## II. Computational details

We performed first principles investigations within the Local Spin-Density Approximation[23] by using the Korringa-Kohn-Rostoker (KKR) band structure method in the Atomic Sphere Approximation (ASA)[24,25], where the partial waves were expanded up to $l_{max}=3$ *(spdf – basis)* inside the atomic spheres. We used the experimental lattice constants of the L2$_1$ lattice structure of the considered alloys1 as listed in Table 1. Since our main goal is to estimate the Néel temperature and the formation of the magnetic order at elevated temperatures, we determined the electronic structure self-consistently in the paramagnetic phase of the considered systems modeled within the Disordered Local Moment (DLM) scheme in the scalar relativistic approximation[26]. Atomic disorder between the Mn and Z sublattices was treated as a random binary alloy, Ru$_2$(Mn$_{1-x}$Z$_x$)(Z$_{1-x}$Mn$_x$) for $0 \le x \le 0.5$ by using the single-site Coherent Potential Approximation (CPA)[25]. As what follows, we shall denote the Mn atoms on the sites of the nominal (original) Mn sublattice by Mn(S), whereas those on the sites of the nominal Z sublattice by Mn anti-sites, Mn(AS).

The magnetic properties of the Ru$_2$MnZ Heusler compounds can well be described by the classical Heisenberg Hamiltonian,

$$H = -\sum_{<i,j>} J_{ij} \vec{e}_i \vec{e}_j \ , \tag{1}$$

where the sum runs over all the different Mn-Mn pairs and $\vec{e}_i$ denotes the unit vector pointing along the magnetic moment of the *i*-th Mn site. In case of chemical disorder between the Mn and Z sublattices, the sum in Eq. (1) also includes sites on the Z sublattice occupied by Mn atoms with probability x. The isotropic exchange interactions, $J_{ij}$, were evaluated from the DLM reference state in the spirit of magnetic force theorem[27] as implemented within the bulk KKR method[28]. The use of the DLM reference state in the calculations of the exchange interaction constants allows for an account of the influence of the thermal magnetic disorder on the electronic structure and on the interatomic exchange interactions, thus, for a more precise estimation of the magnetic ordering temperature (see, e.g. Refs. [22,29,30,31]).

In order to study finite temperature magnetic properties we performed Monte Carlo simulations with the spin-Hamiltonian (1) using 16x16x16 primitive cells of the underlying magnetic fcc lattice (a cluster of 4096 Mn atoms) with periodic boundary conditions. For the simulations of the chemically disordered Ru$_2$MnSi alloys, 12x12x12 cells of both the Mn and Si fcc sublattices, i.e. in total 3456 sites, were randomly filled with Mn atoms according to their partial occupation numbers. Note that in the Monte-Carlo simulations we considered a spin-model with exchange interactions up to the 20$^{th}$ NN shell.

### III. Fully ordered L2$_1$ alloys

First we performed calculations as outlined above for the Ru$_2$MnZ compounds in the fully ordered L2$_1$ structure. The calculated local moments of the Mn atoms, the first three nearest neighbor Mn-Mn exchange interactions and the Néel temperatures obtained from Monte-Carlo simulations are summarized in Table I. The Néel temperatures are compared with the experimental values shown in the last row of Table I. As can be inferred from Table I, the local magnetic moment of Mn in the paramagnetic state is around 3 $\mu_B$ or even higher, indicating a strong localization of the moments as it is usual in Mn-based Heusler alloys. Note, that the variation of the values of the experimental (and calculated) Néel temperatures for the Ru$_2$MnZ series correlate neither with the values of the lattice constant, nor with the size of the local moments of the Mn atoms. Quite obviously, the Néel temperature is determined by the Mn-Mn exchange interactions governed by the actual electronic structure depending on the type of the *p*-element in the Z position.

In ordered Ru$_2$MnZ alloys the magnetic Mn atoms fully occupy one of the four interpenetrating fcc sublattices of the L2$_1$ structure. Thus the AFM ordering occurs on a magnetically frustrated fcc lattice. From earlier experiments[18] it is known that the magnetic structure corresponds to the 2$^{nd}$ kind of AFM order that occurs due to the strong 2$^{nd}$ NN AFM coupling, for which a *p*-atom takes place between the two Mn atoms. Thus, the *p*-elements lead to a strong enhancement of the Mn-Mn interaction via indirect exchange coupling mechanism. The 1$^{st}$ NN direct exchange is ferromagnetic, but it is clearly smaller in magnitude than the 2$^{nd}$ NN AFM coupling. The competition between these two exchange couplings leads to the formation of the 2$^{nd}$ kind of AFM structure on the fcc lattice.

The Monte-Carlo simulations with exchange interactions calculated up to the 20$^{th}$ NN shells result indeed to the 2$^{nd}$ kind of AFM ordering for all the considered alloys. Taking into account only the first three NN couplings in Table I also stabilizes the 2$^{nd}$ kind of AFM structure in full agreement with the ($J_1$, $J_2$, $J_3$) phase diagram of the magnetic fcc lattice as given by Moran-Lopez *et al.*[32]. However, we find an essential difference between our calculated exchange interactions and earlier estimations[19] made on the basis of total energies calculations for a couple of ordered magnetic configurations. This can be understood since in Ref. [19] the authors limited their mapping procedure only to the first three NN interactions and a similar strategy has been pursued in the theoretical analyses of the experimental data in Ref. [18]. In Fig. 2 we show the calculated values of the exchange interactions for distant pairs. Apparently, the 4$^{th}$ NN interactions are rather large for all compounds of the Ru$_2$MnZ series. Moreover, the 3$^{rd}$ NN interactions are very small, even smaller than the 5$^{th}$ and 6$^{th}$ NN couplings. Thus by limiting the mapping of the total energies onto the first three NN interactions, one can make a severe numerical error in the estimated values of these exchange couplings. This nicely illustrates the advantage of the torque method[27] over the direct total energy mapping using a limited number of magnetic configurations in case of metallic magnets with long-range exchange interactions.

One can see from Table I that the calculated Néel temperatures are in excellent agreement with the experiment for the Ru$_2$MnSb and Ru$_2$MnSn compounds, but by about 50 K and 100 K higher than the experimental values for Ru$_2$MnGe and Ru$_2$MnSi, respectively. The most possible reason for this disagreement is the partial chemical disorder within the Mn and Z sublattices observed in the single crystals of Ru$_2$MnSi, whereas an almost perfect L2$_1$ order has been reported for Ru$_2$MnSb and Ru$_2$MnSn[18]. We will investigate the effect of partial chemical disorder on the Néel temerature after discussing the frustration effects on the magnetic correlation functions in the next section.

### IV. Effects of magnetic frustration

The magnetic order in Ru$_2$MnZ alloys can be understood on the basis of four interpenetrating simple cubic lattices constituting an fcc lattice, each of them possessing a checkerboard AFM order. The mutual orientations of the sublattice moments are frustrated and this frustration is even not removed by considering more distant interactions beyond the third NN shell. In this Section we illustrate the manifestation of frustration effects in the finite temperature spin-spin correlation functions. As an example we take the Ru$_2$MnSb compound, which exibits a very high degree of chemical order[18]. The spin-spin correlation function for the $n^{th}$ nearest neighbour shell is defined as,

$$c(n) = \frac{1}{N}\sum_i \frac{1}{N_n}\sum_{\vec{R}_n} \langle \vec{e}_{\vec{R}_i} \vec{e}_{\vec{R}_i+\vec{R}_n} \rangle \,, \qquad (2)$$

where the first sum runs over $N$ translation vectors of the fcc lattice, $\vec{R}_i$, the second sum is taken over the $N_n$ translation vectors, $\vec{R}_n$, spanning the $n^{th}$ shell, and $\langle \rangle$ stands for the statistical average. Quite obviously, these correlation functions provide information on the magnetic short range order in the system.

It can be easily shown that, in the ordered 2$^{nd}$ kind of AFM structure, the spin-spin correlation function for the 2$^{nd}$ and 4$^{th}$ NN shells takes the values -1 and +1, respectively, since all the corresponding neighbors are uniformly magnetized antiparallel or parallel with respect to the atom at the arbitrarily chosen center position. In the upper panel of Fig. 3 the temperature dependence of the spin-spin correlation functions is displayed for the ordered Ru$_2$MnSb alloy. At low temperatures, the functions $c(2)$ and $c(4)$ reach indeed values close to -1 and +1, respectively, whereas $c(n)$, for $n = 1, 3, 5,$ and $6$ approach to zero. The magnitudes of $c(2)$ and $c(4)$ monotonously decrease as the temperature increases. The inflection point of the curves indicates the ordering temperature, $T_N = 180\ K$ in good agreement with experimental value of $195\ K$.

A specific feature of the 2$^{nd}$ kind of AFM structure on the fcc lattice is that the 1$^{st}$, 3$^{rd}$ and 5$^{th}$ shell correlation functions (and all beyond the 5$^{th}$ shell) vanish as the temperature approaches to zero. This happens since the respective shells contain equal number of sites with opposite magnetizations. However, since the 1$^{st}$ NN interaction in Ru$_2$MnSb is quite strong relative to the 2$^{nd}$ and 3$^{rd}$ NN interactions, a ferromagnetic short-range order likely develops in the paramagnetic phase. This is illustrated in the lower panel of Fig. 3 in terms of the temperature dependence of the spin-spin correlation functions, $c(1)$ and $c(5)$. As can be seen, these correlation functions take a finite value well above the critical temperature in the paramagnetic phase and, when the system is cooled down, they even gradually increase. Upon the onset of AFM order below the critical temperature, the respective short range order rapidly decreases and vanishes at zero temperature. A detailed discussion of the magnetic short range order in the paramagnetic phase can be found in Ref. [33], where a similar analysis of the spin-spin correlation functions was performed for the AFM GdPtBi half-Heusler alloy.

## IV. Effects of chemical disorder on the magnetism in Ru$_2$MnSi

In order to properly describe the magnetism in Ru$_2$MnSi we took into account the partial chemical disorder between the Mn and Si sublattices, i.e. the presence of some fraction of anti-site (AS) Mn atoms on the Si sublattice (or, other way around, the presence of Si(AS) on the Mn sublattice). In this section we present the results of DLM calculations for partially disordered Ru$_2$(Mn$_{1-x}$Si$_x$)(Si$_{1-x}$Mn$_x$) alloys along the path from the L2$_1$ to the B2 phase ($0 \leq x \leq 0.5$). The calculated exchange interactions for close neighbors are shown in Fig. 4 as a function of the concentration $x$. Concerning the Mn–Mn pairs on the nominal Mn sublattice, the 2$^{nd}$ NN AFM interaction is slightly reduced, whereas the 1$^{st}$ and 4$^{th}$ NN FM interactions rapidly decrease with increasing disorder and the 1$^{st}$ NN interaction even changes sign at about x=0.3. Noteworthy, the 2$^{nd}$ NN Mn(S)–Mn(AS) interaction is strongly ferromagnetic, while the leading Mn(AS)–Mn(AS) interactions are antiferromagnetic. Clearly, the interactions for the respective Mn-Mn pairs at the proper sites and at the anti-sites become identical in the B2 phase ($x = 0.5$).

We performed Monte-Carlo simulations, where the magnetic sites were randomly distributed over the the combined Mn-Si sublattices with the prescribed concentrations. The simulated Néel temperatures are shown in Fig. 5. One can immediately see that the chemical disorder significantly decreases the Néel temperature in the Ru$_2$MnSi alloys. The calculated Néel temperature agrees well with the experimental one for about x(Si)=0.1, which is consistent with a weak disorder found in the experimental samples[18]. In addition, we can conclude that the Néel temperature of Ru$_2$MnSi can be increased above the room temperature by producing the samples with better L2$_1$ order.

Note, that in case of partially ordered alloys the underlying magnetic sublattice is simple cubic. As the number of Mn atoms on the Si sublattice increases, the strong FM Mn(S)–Mn(AS) interactions start to play a dominant role and finally the simulations predict the formations of a complex random AFM structure being quite different from the initial 2$^{nd}$ kind of AFM ordering on the fcc lattice. This AFM phase stabilizes on a strongly disordered simple cubic (sc) lattice and it doesn't correspond to any type of AFM collinear ordering on a chemically ordered sc lattice. The spin structure is periodic along the [1 1 1] direction with a wave vector of $\vec{q}$ = [½ ½ ½]. This periodicity is consistent with a maximum of the Fourier transform, $J(\vec{q})$, of the calculated exchange interactions for the B2 phase assuming that all sites of the underlying sc lattice is populated by Mn atoms. Our simulations thus predict the change of the type of AFM ordering on the path from the L2$_1$ to the B2 phase. In Fig. 5 the approximate concentration where this phase transition occurs is marked by a vertical line. It is, however, not clear whether this magnetic phase can be observed in the experiment since it requires a very high degree of the chemical disorder in the Mn-Si sublattices.

## V. Conclusions.

In terms of combined first principles calculations and Monte Carlo simulations, we have shown that the antiferromagnetic structure of Ru$_2$MnZ (Z=Sn, Sb, Ge, Si) full-Heusler alloys is determined by a strong second nearest neighbor antiferromagnetic coupling between well localized Mn moments. This interaction is mediated by the *p*-atom positioned between the interacting pair of Mn atoms. The calculations also evidence that a strong ferromagnetic 4$^{th}$ NN coupling strengthens the stability of 2$^{nd}$ kind of AFM structure. The calculated Néel temperatures for fully ordered L2$_1$ structures are in excellent agreement with experiment for the alloys exhibiting very high degree of chemical ordering (Z = Sn and Sb). For Ru$_2$MnSi, where a moderate disorder in the Mn-Si sublattice is determined experimentally, we demonstrated that the chemical disorder significantly reduces the critical temperature and we found that the experimental Néel temperature can be reproduced with about 20% of Mn anti-site atoms on the Si sublattice. This observation might be of great importance for applications, since the Néel temperature of Ru$_2$MnSi can be pushed well above the room temperature by lowering the chemical order in the samples. Moreover, for alloys close to the B2 order we predicted the appearance of a new AFM phase corresponding to the wave vector $\vec{q} = [½\ ½\ ½]$.

The research leading to these results has received funding from the European Union Seventh Framework Programme (FP7/2007-2013) under grant agreement no. NMP3-SL-2013-604398. Financial support was also provided in the frames of the TÁMOP-4.2.4. A/2-11-1-2012-0001 ''Hungarian National Excellence Program'' co-financed by the European Union and the European Social Fund, as well as by the Hungarian Scientific Research Fund under contract OTKA K84078.

**Table I:** Calculated local magnetic moments of Mn atoms ($m_{Mn}$), first three nearest neighbor exchange interactions between the Mn atoms ($J_{1NN}$, $J_{2NN}$, and $J_{3NN}$), and Néel temperatures ($T_N^{calc}$) for ordered Ru$_2$MnX compounds. The experimental lattice constant ($a$) and Néel temperatures ($T_N^{exp}$) are taken from Ref. [18]. A representative relative position vector of the corresponding Mn-Mn pair is given in units of $a$ below the labels of the exchange interactions.

|  | $a$ (Å) | $m_{Mn}$ ($\mu_B$) | $J_{1NN}$ (mRy) [½ ½ 0] | $J_{2NN}$ (mRy) [1 0 0] | $J_{3NN}$ (mRy) [1 ½ ½] | $T_N^{calc}$ (K) | $T_N^{exp}$ (K) |
|---|---|---|---|---|---|---|---|
| Ru$_2$MnSn | 6.217 | 3.26 | 0.11281 | -0.49540 | -0.01640 | 320 | 296 |
| Ru$_2$MnSb | 6.200 | 3.56 | 0.14834 | -0.20500 | 0.00646 | 180 | 195 |
| Ru$_2$MnGe | 5.985 | 3.04 | 0.17679 | -0.48553 | -0.01742 | 365 | 316 |
| Ru$_2$MnSi | 5.887 | 2.95 | 0.17646 | -0.52136 | -0.02093 | 415 | 313 |

**Figure captions**

**Figure 1 (color online):** The L2$_1$ crystal structure of Ru$_2$MnZ alloys (upper panel) and the 2$^{nd}$ kind of AFM order on the fcc sublattice of Mn atoms (lower panel).

**Figure 2 (color online):** Calculated Mn-Mn exchange interactions, $J_{ij}$, in fully ordered Ru$_2$MnZ (Z=Ge, Sn, Si, Sb) compounds as a function of the distance, $d$, measured in units of the lattice constant, $a$. The relative position vectors of the pairs are labelled explicitely for the first four nearest neighbors.

**Figure 3 (color online):** Calculated temperature dependence of the spin-spin correlation functions, Eq. (2), for the selected neighbor shells in Ru$_2$MnSb. Note the difference in the vertical scale of the two panels.

**Figure 4(color online):** Calculated Mn-Mn exchange interactions in partially ordered Ru$_2$MnSi alloy as a function of the Si concentration in the Mn sublattice. Upper panel: interactions between the Mn atoms on the nominal Mn sublattice. Lower panel: Mn(S)-Mn(AS) interactions (combined symbols) and Mn(AS)-Mn(AS) interactions (open symbols).

**Figure 5 (color online):** Simulated Néel temperature of partially ordered $Ru_2MnSi$ alloy as a function of the Si concentration in the Mn sublattice. The approximate concentration where a transition in the AFM ordering occurs (see text) is marked by vertical line.

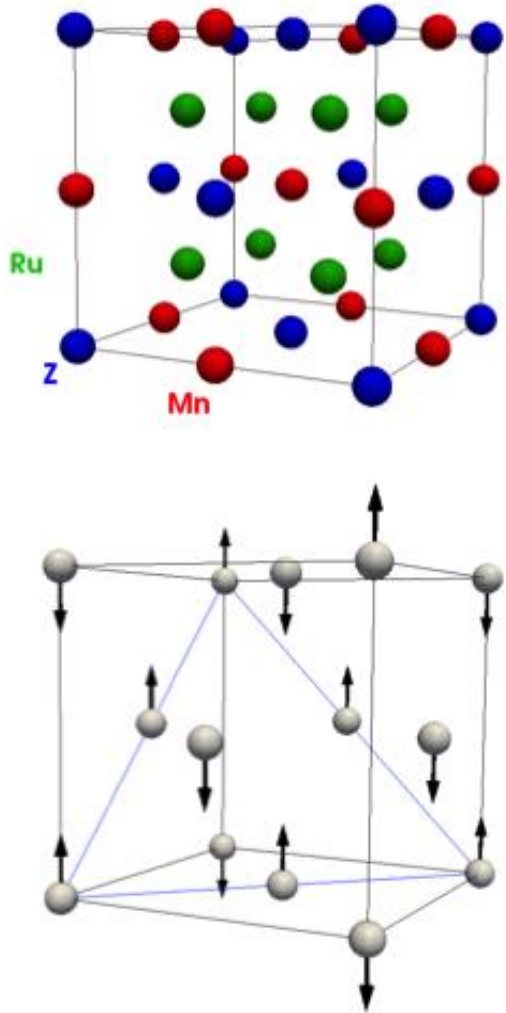

**Figure1_Khmelevskyi**

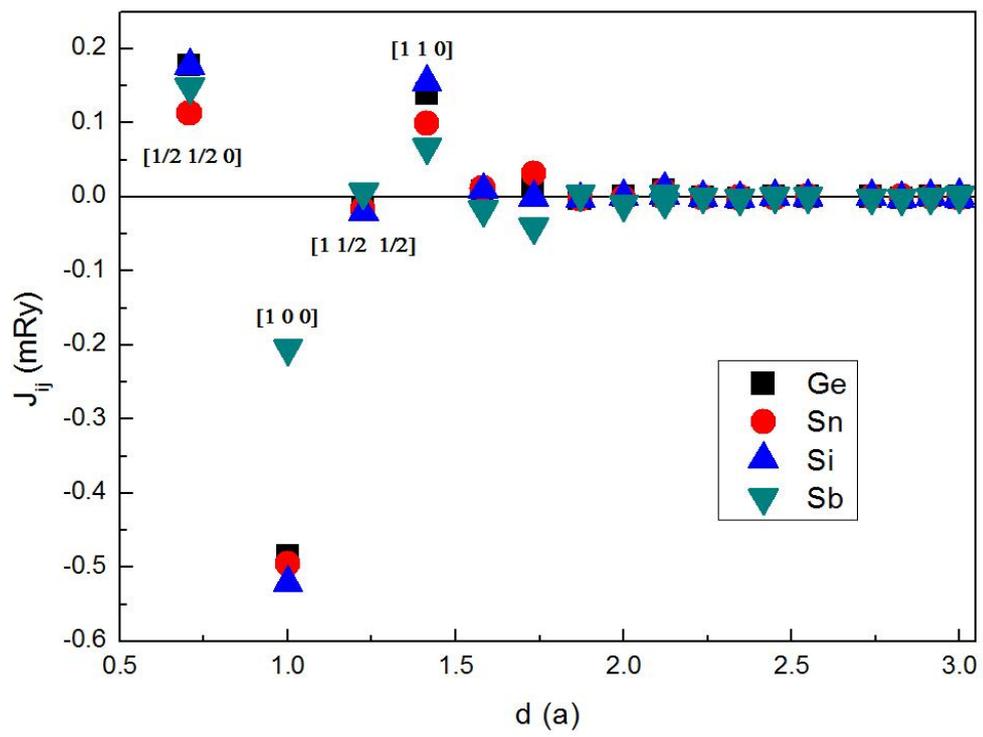

Figure2_Khmelevskyi

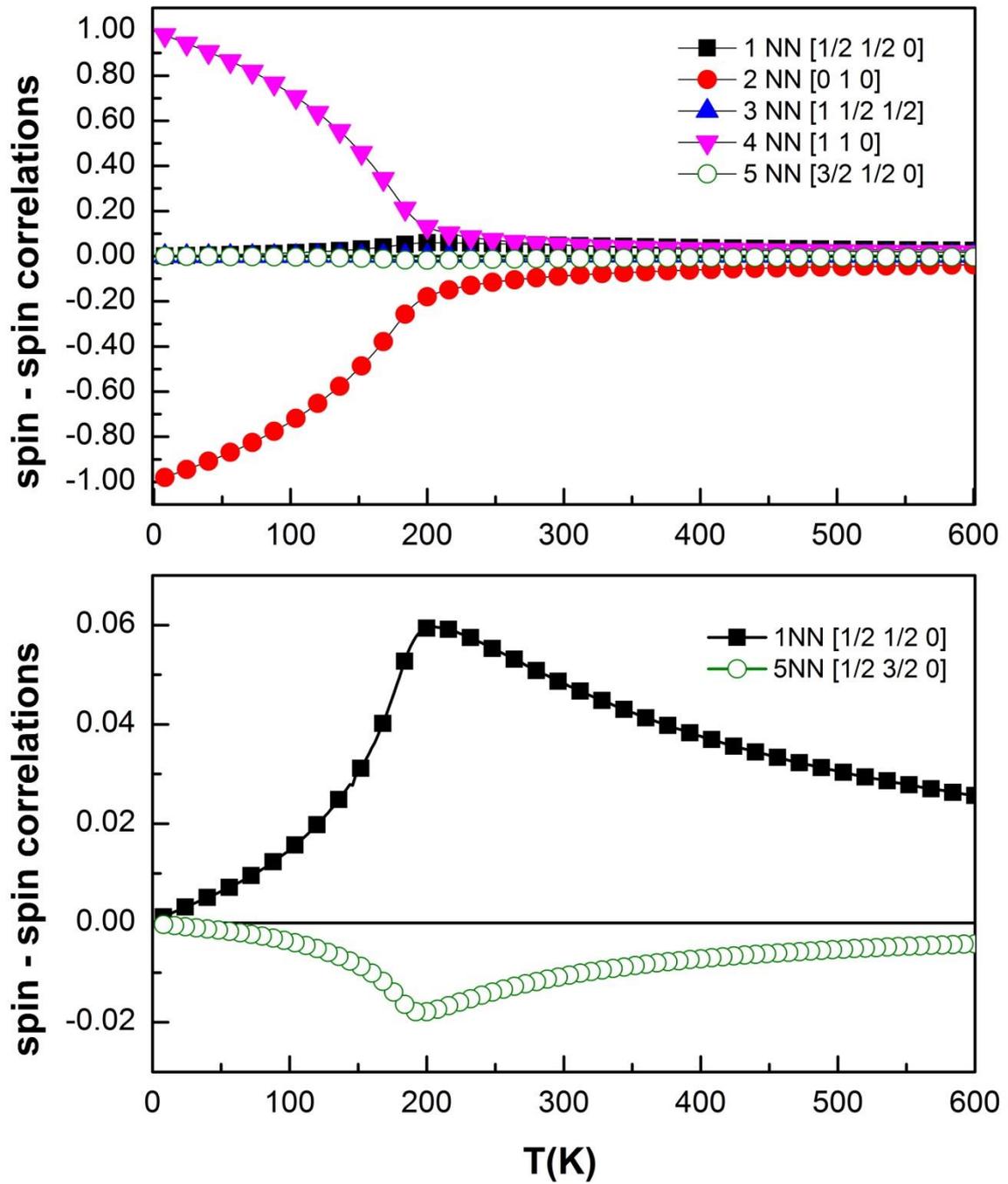

**Figure3_Khmelevskyi**

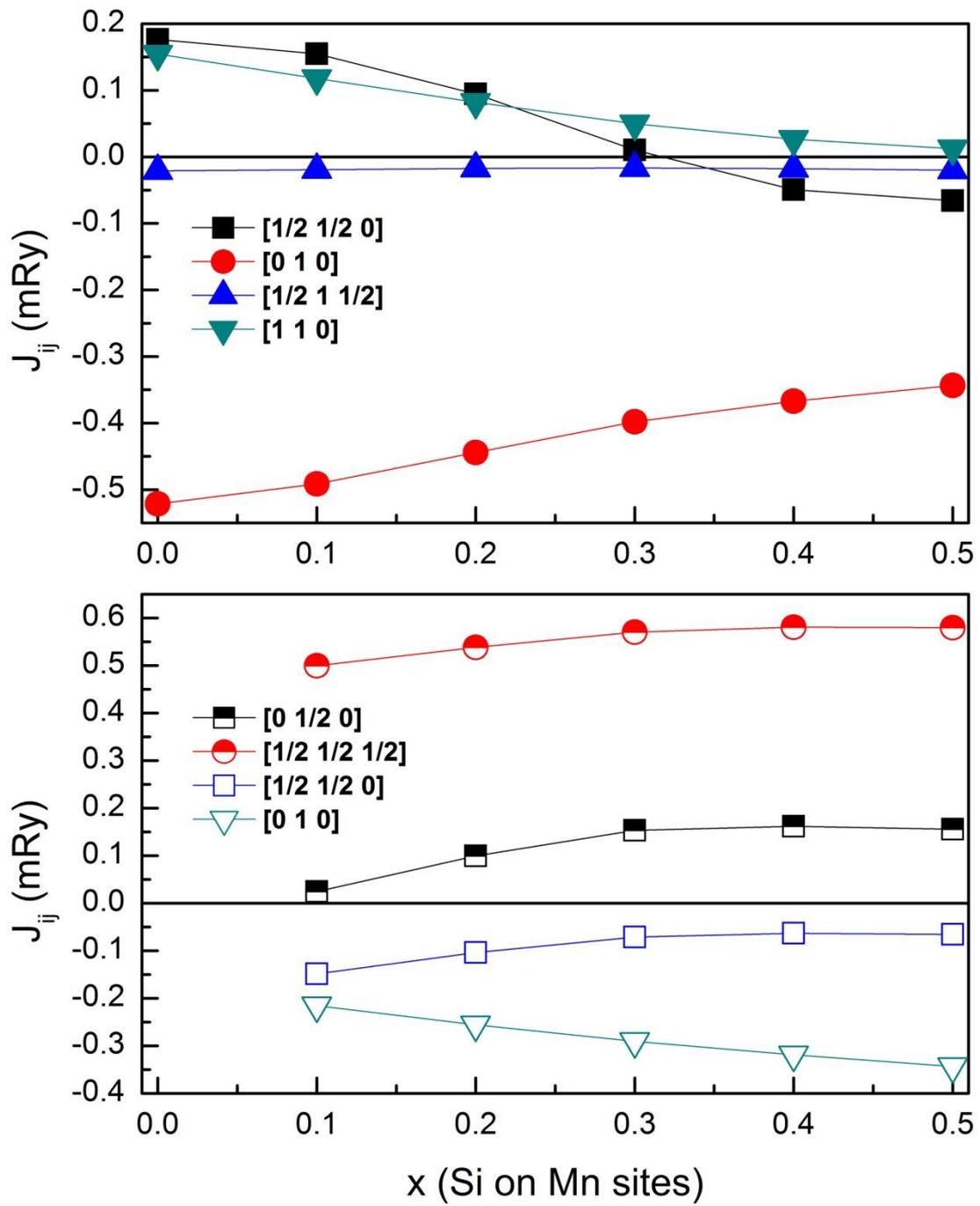

**Figure4_Khmelevskyi**

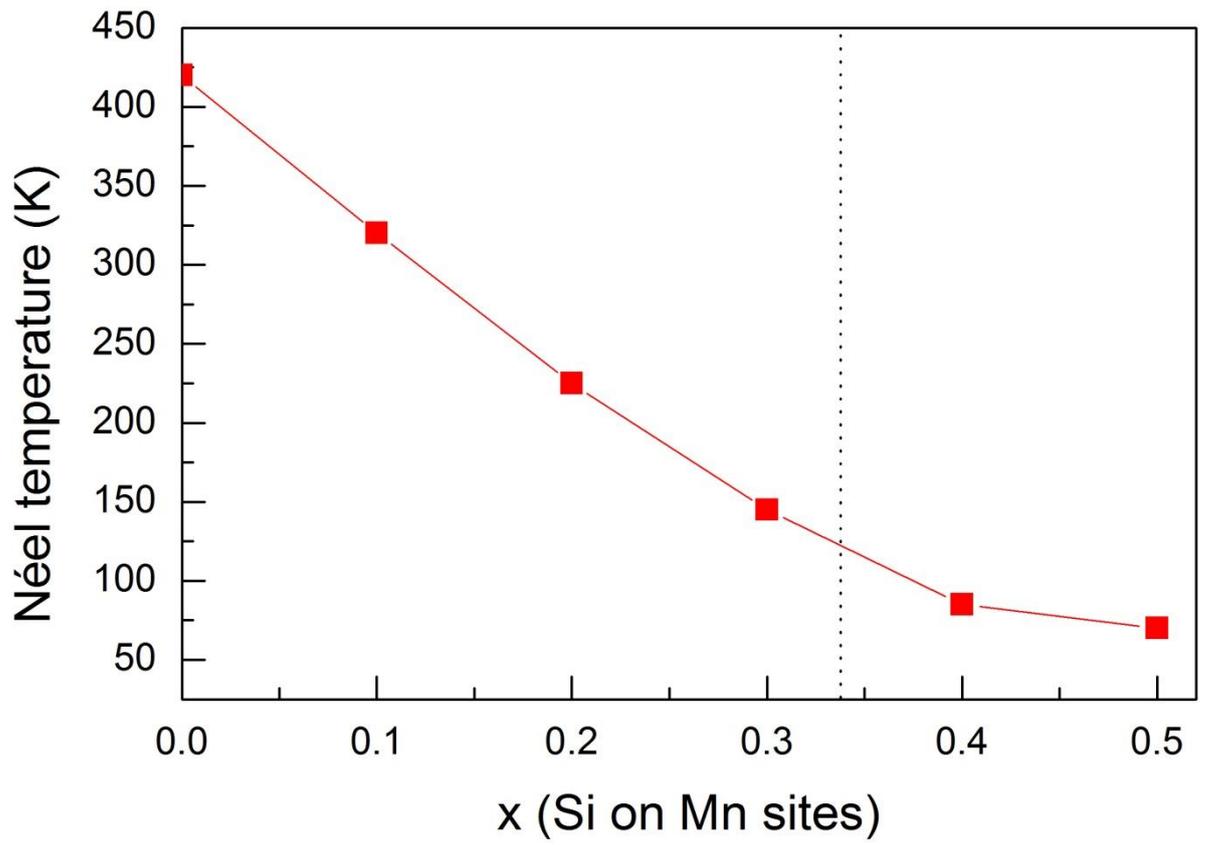

**Figure5_Khmelevskyi**